\documentclass[preprint]{elsarticle}
\usepackage{bm}
\usepackage[cp1251]{inputenc}
\usepackage[english]{babel}
\usepackage{amsfonts,amssymb,amsmath}
\usepackage{amsmath,graphicx}
\usepackage{graphicx,pifont} 
\usepackage{epsfig}
\title{Towards the microscopic foundation of the zeroth law of thermodynamics}
\author[ayz]{A.Yu.~Zakharov}
\ead{Anatoly.Zakharov@novsu.ru}
\author[maz]{M.A.~Zakharov}
\ead{Maxim.Zakharov@novsu.ru}
\address{Yaroslav-the-Wise Novgorod State University, Veliky Novgorod, 173003, Russia}
\begin{document}
	\begin{abstract}
The dynamics of free vibrations of a chain of atoms is investigated taking into account the retardation of interactions. It is shown that all oscillations of the circuit are damped. The dynamics of forced vibrations of this chain of atoms is investigated. It is shown that, regardless of the initial conditions, the system passes into a stationary state of dynamic equilibrium with an external field, which depends both on the properties of the system and on the parameters of the external field. A non-statistical dynamic mechanism of the process of irreversible establishment of the state of thermodynamic equilibrium in both many-body and few-body systems is proposed.
\end{abstract}
	
	\begin{keyword}
		zeroth law of thermodynamics; crystal dynamics; retarded interactions; irreversibility; equilibration dynamics 
		
		MSC classes: 80A05 \ 80A10 \ 37K60 \ 37L60 \ 70H40
		
		\PACS: 05.20.-y \sep 63.10.+a \sep 05.70.Ln \sep 45.50.Jf
	\end{keyword}
	
	\maketitle

\section{Introduction}
The goal of statistical mechanics~\cite{Gibbs} is the microscopic foundation the phenomenological laws of thermodynamics. Despite the outstanding achievements of statistical mechanics in calculating and explaining the properties of matter, a number of fundamental problems of thermodynamics remain open.
First of all, this is the problem of correct foundation the zeroth principle of thermodynamics, i.e. statements about the existence of a state of thermodynamic equilibrium in many-particle systems. According to this postulate, any isolated system, consisting of a very large number of atoms, over time passes into a state of thermodynamic equilibrium, in which all macroscopic variables reach constant values~\cite{Guggenheim,Uhlenbeck,Lieb1,Lieb2,Brown1}.

However, the zeroth principle of thermodynamics in the framework of statistical mechanics is postulated in the same way as the application of the probability concept to describe the evolution of dynamical systems in the absence of clear sources of stochastization~\cite {Kubo1}. In reality, the combination of the principles of deterministic and reversible classical mechanics with the concept of probability leads to an internal contradiction, the satisfactory settlement of which is not yet available~\cite{Landau1, Strien}.

Of the many paradoxes that arise in attempts to substantiate thermodynamics within the framework of combining classical mechanics with the concept of probability, the Kac ring model~\cite{Kac1, Kac2} is especially interesting. This purely mechanical model admits an exact analytical solution that is reversible in time and satisfies the Poincar\'e recurrence theorem. However, the application of very plausible probabilistic hypotheses like the assumption of ``molecular chaos'' to the ring model leads to irreversible dynamics of this model, which is in clear contradiction with the exact solution~\footnote{Note that the chaotic nature of trajectories in classical mechanics, caused by the instability of solutions of dynamic equations with respect to variations in the initial conditions (for example, the Sinai billiards), has no direct relation to the justification of the zero principle of thermodynamics. The point is that the irreversibility of dynamics is a necessary but not a sufficient condition for the establishment of thermodynamic equilibrium in the system.}. 
Therefore, the origin of macroscopic irreversibility and the nature of the establishment of thermodynamic equilibrium remain unknown~\cite{Landau1}.

In 1909, shortly after the creation of statistical mechanics, the paradoxical work of Ritz and Einstein~\cite{Ritz} was published. The authors expressed mutually exclusive points of view on the nature of irreversibility: ``Ritz considers the restriction to the form of the retarded potentials as one of the roots of the second law [of thermodynamics] while Einstein believes that irreversibility depends exclusively upon reasons of probability'' (``Ritz - Einstein Agreement of Disagreement''~\cite{Fritzius}).

The aim of this work is to present a possible non-statistical mechanism of the process of establishing of thermodynamic equilibrium in a system of particle. To illustrate this mechanism, the dynamics of a one-dimensional chain of atoms with retarded interactions between particles is studied.

\section{Free oscillations of a chain of particles with retarded interactions}

Consider a one-dimensional system of identical particles interacting with each other, whose stable equilibrium positions form an ideal lattice with the Born-von~Karman boundary conditions~\cite{Born1}
\begin{equation}\label{x-n}
	x_{n}^{\left( 0\right) } = n\, a\quad  \left(a = \mathrm{const},\quad  x_{n+N}^{\left( 0\right) } = x_{n}^{\left( 0\right) }, \quad  1 \leq n \leq N\right),
\end{equation}
where $ a $ is the distance between the nearest neighbors of the lattice.

The local value of the potential of the field created by all particles at the site $ x_{n}^{\left (0 \right)} $ for instantaneous interactions has the form
\begin{equation}\label{phi-0}
	\varphi\left( x_{n}^{\left(0 \right) } \right) = \sum_{\stackrel{n'}{\left( n'\not= n\right) }}   \, v\left(x_{n}^{\left(0 \right) }  - x_{n'}^{\left(0 \right) }\right), 
\end{equation}
where  $ v\left(x \right)  $ is the energy of pair interaction of two atoms located at a distance $ x $ from each other.

In this case, the dynamics of the system in the harmonic approximation is described by the equations~\cite{Born4}
\begin{equation}\label{instant}
	m \ddot{U}_{n}\left( t\right) = \sum_{n'>0} v'' \left( n' \right)\, \left[ U_{n-n'}\left( t\right)  -2 U_{n}\left( t\right)   +  U_{n+n'}\left( t\right)\right],  
\end{equation}
where $ m $ is the mass of the atom, $ {v}''\left(n \right) $~is the second derivative of the function $ v\left(x \right) $ at $ x = na $, $ U_ {n} \left(t \right) $~is displacement of the $ n $-th particle from its equilibrium position:
\begin{equation}\label{U-n}
	{U}_{n}\left( t\right) = x_{n}\left( t\right)  - x_{n}^{\left( 0\right) }\left( t\right), \quad \left| {U}_{n}\left( t\right) \right| \ll a. 
\end{equation}

It is known that solutions of the equations of crystal lattice dynamics with \textit{instantaneous interactions} in the harmonic approximation lead to the concept of phonons. The dispersion law of these quasiparticles depends on the crystal structure and interatomic potentials.

However, real interactions between particles always have the property of retardation due to the finite speed of propagation of interactions.
This property leads to a radical change in the dynamics even in the simplest case of a two-body problem, including the irreversible behavior of the~\cite{Synge, Zakharov2019} system.

To take into account the effect of retardation of interactions between particles of a one-dimensional lattice in the equation~\eqref {instant}, we make the replacement
\begin{equation}\label{tau1}
	U_{n\pm n'}\left( t\right) \longrightarrow U_{n\pm n'}\left( t - \tau\left(n'a  \right)  \right),
\end{equation}
where $ \tau \left (n'a \right) $~is the retardation time of interaction between points located at a distance $ n'a $ from each other.

By virtue of the condition~\eqref{U-n}, we assume that the retardation of interactions between each pair of particles depends only on the equilibrium distances between them. Since the retardation of interactions between points are proportional to the distances between them, we put
\begin{equation}\label{tau-n}
	\tau\left(n'a  \right) = \frac{an'}{c}=\tau_{1}\,n',
\end{equation}
where $ c $ is the speed of propagation of interactions between particles, that is, the speed of light, $ \tau_{1} $~is the retardation time of the interaction between the nearest neighbors of the lattice.

Thus, the equations of the dynamics of a one-dimensional chain of interacting particles in the harmonic approximation, taking into account the retardation of interactions, have the following form
\begin{equation}\label{1D-retard}
	\left\lbrace 
	\begin{array}{l}
		{\displaystyle  m \ddot{U}_{n}\left( t\right) = \sum_{n'>0} v'' \left( n' \right)\, \left[ U_{n-n'}\left(  t\ - n' \tau_{1}  \right)  -2 U_{n}\left( t\right)   +  U_{n+n'}\left(  t\ - n' \tau_{1} \right) \right];  }\\
		{\displaystyle U_{n+N}\left(t \right) = U_{n}\left(t \right) .}
	\end{array}
	\right. 
\end{equation}

We will seek a solution to this system of equations in the form
\begin{equation}\label{u-n(t)}
	U_{n}\left( t\right) = Q_{k}\left( t\right) \, e^{i\,kna},
\end{equation}
where $ Q_{k} \left (t \right) $ are normal coordinates. It follows from the Born-von Karman boundary conditions that
\begin{equation}\label{kaN}
	k = 2\pi\frac{s}{aN}
\end{equation}
($ s $ is an arbitrary integer)
and
\begin{equation}\label{k}
	\frac{\pi}{a} \leq	k < \frac{\pi}{a}.
\end{equation}
Substituting~\eqref{u-n(t)} into the equations~\eqref{1D-retard}, we get
\begin{equation}\label{u-tau}
	m\ddot{Q}_{k}\left( t\right) -\sum_{n'>0}v''\left( n'\right) \left[Q_{k} \left( t-n'\tau_{1}\right)\, e^{-ikan'} -2\, Q_{k}\left( t\right) +  Q_{k} \left( t-n'\tau_{1}\right)\, e^{ikan'}\right] = 0.
\end{equation} 

Let us substitute
\begin{equation}\label{1D-char}
	Q_{k}\left(t \right) = Q\, e^{-i\omega t}
\end{equation}
and obtain an equation for the dispersion law $ \omega \left(k \right) $:
\begin{equation}\label{omega}
	m\omega^{2} - 2 \sum_{n'>0} v'' \left( n' \right) \left[1 - e^{i\omega\tau_{1} n'} \cos\left(kan' \right) \right] = 0. 
\end{equation}
This equation in the general case (that is, at $ \tau_{1} \neq 0 $) is transcendental, and the set of its roots is infinite.

Taking into account that
\begin{equation}\label{Omega-tau}
	\omega\tau_{1} \sim v/c \ll 1
\end{equation} 
($ v $~is the characteristic velocity of motion of atoms), we select from the infinite set of roots of the characteristic equation a finite number of ``actual'' roots. In this case, the transcendental characteristic equation~\eqref{omega} is reduced to a quadratic equation for the ``actual'' roots.
\begin{equation}\label{omega1}
	\omega^{2}	+ 2i \omega \frac{\tau_{1}}{m} \sum_{n'>0} n'\, v''\left( n'\right) \cos\left( kan' \right)  - \frac{4}{m} \sum_{n'>0}\, v''\left(n' \right) \sin^{2}\left(\frac{kan'}{2} \right)  =0.
\end{equation}
Thus, the `` actual '' roots of the characteristic equation have the form
\begin{equation}\label{roots}
	\omega_{1,2}\left( k \right) = \pm \sqrt{\frac{4}{m} \sum_{n'>0}\, v''\left(n' \right) \sin^{2}\left( \frac{kan'}{2}\right)  } -i \frac{\tau_{1}}{m} \sum_{n'>0}\, n' \, v''\left( n'\right) \, \cos\left( kan'\right). 
\end{equation}

Note that the imaginary parts of all ``actual'' roots of the characteristic equation are negative; therefore, substitution of~\eqref{roots} into~\eqref{1D-char} leads to the conclusion that all oscillations in the system are damped.

Compare the real and imaginary parts of $ \omega_{1,2}\left( k \right) $
\begin{equation}\label{Re-Im}
	\left| \frac{\mathrm{Im}\ \omega\left( k\right) }{\mathrm{Re}\ \omega\left( k\right) } \right|  = \frac{\displaystyle \frac{\tau_{1}}{m} \sum_{n'>0} n'\, v''\left( n'\right) \cos\left( kan'\right) }{\displaystyle \sqrt{\frac{4}{m} \sum_{n'>0}\, v''\left(n' \right) \sin^{2}\left( \frac{kan'}{2}\right)  } }.
\end{equation}
The asymptotics of this ratio in a vicinity of the point $ k = 0 $ has the form
\begin{equation}\label{asymp}
	\left| 	\frac{\mathrm{Im}\ \omega\left( k\right) }{\mathrm{Re}\ \omega\left( k\right) }\right| 	\approx \frac{\displaystyle  \sum_{n'>0} n'\, v''\left( n'\right) }{\displaystyle \sqrt{\sum_{n'>0}\, \left( n'\right)^{2} v''\left(n' \right)  } }\, \frac{\tau_{1}}{\sqrt{m}\ \left| a\,k\right|} \to \infty.
\end{equation}
Therefore, the long-wavelength oscillations ($ \left| ak \right| \ll 1 $) are damped most rapidly. When $ t \to \infty $, all oscillations in the chain stop.

\section{Dynamics of forced oscillations of a chain with retarded interactions}

Consider the problem of the dynamics of a one-dimensional atomic chain immersed in an alternating external force field. We denote the external force acting on the normal coordinate $ Q_{k} \left(t \right) $ by $ f_{k} \left(t \right) $. Then the equations of the dynamics of the system have the form:
\begin{equation}\label{extern-f}
	m\ddot{Q}_{k}\left( t\right) + 2\sum_{n'>0}v''\left( n'\right) \left[Q_{k} \left( t \right) - Q_{k} \left( t-n'\tau_{1}\right) \, \cos \left( {kan'}\right)  \right] = f\left( t\right).
\end{equation}
In the first order in the retardation of interactions, we have
\begin{equation}\label{ext-1}
	{\ddot{Q}_{k}\left( t\right) + 2 \Gamma\left( k\right)  \dot{Q}_{k}\left(t \right) + \Omega^{2}\left( k\right) Q_{k}\left( t\right)  = \frac{f\left( t\right)}{m}, }
\end{equation}
where 
\begin{equation}\label{Gamma}
	{\displaystyle  \Gamma\left( k\right) = \frac{\tau_{1}}{m} \sum_{n'>0} v'' \left( n' \right)\,n'\, \cos\left(kan' \right)  }\\
\end{equation}
and
\begin{equation}\label{Omega}
	{\displaystyle \Omega^{2}\left( k\right) = \frac{4}{m} \sum_{n'>0} v'' \left( n' \right)\,  \sin^{2}\left(\frac{kan'}{2} \right). }
\end{equation}

Because of the term $ \dot {Q}_{k} \left(t \right) $, the equation~\eqref{ext-1} has the same form as the equations of forced oscillations under friction. However, the origin of this term is not due to chaotic interaction between particles, as is the case in the framework of stochastic kinetic equations, but to a purely relativistic phenomenon -- the retardation in interactions between particles associated with the existence of a limiting transfer rate of interactions.

The general solution of the linear inhomogeneous system of differential equations~\eqref{ext-1} is the sum of the general solution of the corresponding homogeneous system of equations (which, as shown in the previous section, tends to zero as $ t \to \infty $) and the particular solution of the inhomogeneous system equations. Therefore, over time, stationary oscillations are established in the system, determined by the characteristics of the external field.

\section{Asymptotics of the dynamics of a chain of atoms and thermodynamic equilibrium}

We represent the external force in the form of the Fourier expansion
\begin{equation}\label{Four}
	f\left(t \right) = \sum_{s} C_{s}\, e^{i\gamma_{s}t}.
\end{equation}
Then the solution of the equations~\eqref{ext-1} has the form~\cite {Landau2}
\begin{equation}\label{Four1}
	Q_{k}\left( t\right) = \sum_{s}	\frac{C_{s} }{ m \sqrt{\left[\Omega^{2}\left( k\right) - \gamma_{s}^{2} \right]^{2} + 4 \Gamma^{2} \left( k\right) \gamma_{s}^{2} } }\, e^{i\gamma_{s}t + i \delta_{s}\left( k\right) }, 
\end{equation}
where
\begin{equation}\label{delta}
	\tan\delta_{s}\left( k\right) =\frac{2\Gamma\left( k\right) \gamma_{s}}{\gamma_{s}^{2} - \Omega^{2}\left( k\right) }.	
\end{equation}

Thus, in the limit $ t \to \infty $, the system goes over to a stationary state, which is in a dynamic equilibrium with an external field.

\section{Discussion}

Within the framework of pre-relativistic physics, the essence of the potential energy of interactions between particles remained a kind of `` thing in itself ''. This is some function that depends on the instantaneous configuration of the system and has a hidden origin. In this regard, it is appropriate to note one of the first attempts to find a mechanical interpretation of the interaction of distant bodies: in the outstanding treatise~\cite{Hertz} Heinrich Hertz proved that the potential energy of interacting bodies is mathematically equivalent to the kinetic energy of hidden particles.

In the framework of relativistic physics, instantaneous interaction of particles distant from each other is impossible. The potential energy of a system of interacting particles, depending on their instantaneous positions, does not exist~\cite{Zakharov2019a}. Therefore, a description of the dynamics of a system of atoms on the basis of the Hamiltonian of a system of particles is possible only in the nonrelativistic approximation.
Instead of Hertz's `` hidden bodies '', the field acts as a mediator in particle interactions. Therefore, a correct description of the dynamics of a system of particles interacting through the field must take into account the dynamics of the field. In particular, for the classical system of point charged particles, such a complete system of equations consists of the equations of the relativistic dynamics of particles and the Maxwell equations for the electromagnetic field.

In the paper~\cite{Zakharov2020a}, a classical relativistic dynamic theory of a system of point charged particles interacting through a generated electromagnetic field in the absence of external fields is constructed. The relativistic dynamics of such a system is described in terms of microscopic (i.e., not averaged) distribution functions. Within the framework of this approach, an exact analytical elimination of field variables is performed and a finite closed system of differential-functional equations of retarded type with respect to microscopic distribution functions of particles is obtained. This system of equations is not invariant with respect to time reversal, since the advanced electromagnetic fields were omitted due to the principle of causality~\footnote{It should be noted that the retardation of interactions leads not only to the phenomenon of irreversibility, but also manifests itself in a qualitative change in the spectrum of elementary excitations in crystals. In particular, as shown in~\cite{Lerose}, it is responsible for the splitting of the optical branches in the dispersion curves of ionic crystals near the electromagnetic dispersion line  $ \omega = ck $.}.


Thus, the phenomenon of irreversibility of the dynamics of a system of particles is generated by two reasons:
\begin{itemize}
	\item the field nature of the interaction between particles;
	\item the principle of causality.
\end{itemize}

However, only the irreversibility of dynamics is not enough for a microscopic explanation of the existence of thermodynamic equilibrium in the system.

The main results of this work are as follows.
\begin{enumerate}
\item It has been established that the relativistic effect of retardation in interactions between the particles forming an one-dimensional chain leads to complete freezing of oscillations, i.e. to the transition of the system to its ground state. The kinetic energy of the atoms forming this chain is completely and irreversibly transformed into the energy of the field, through which the atoms interact.
\item It has been showed that system, when immersed in a one-dimensional chain of particles with delayed interactions between them in an alternating external field, ``forgets'' its initial state and passes into a stationary state, which depends both on the properties of the system and on the characteristics of the external field.
\item It has been established that the relativistic effect of retardation in interactions between atoms and desynchronization of oscillations associated with retardation is a mechanism by which a dynamic equilibrium is established between a system of particles and an external field.
\item It has been established that the field through which the interaction between particles occurs plays the role of an inexhaustible thermodynamic reservoir. This reservoir with an infinite number of degrees of freedom ensures the realization of the zero principle of thermodynamics, that is, the irreversible process of establishing of thermodynamic equilibrium both in many-body systems and in few-body systems.
\end{enumerate}

\section{Conclusion}

Thus, the behavior of the system under study in the framework of relativistic physics is radically different from the behavior of the same system in the framework of pre-relativistic physics, in which an isolated chain of atoms oscillates \textit {eternally}. In relativistic physics, both irreversibility and a state of dynamic equilibrium exist. There is neither one nor the other in pre-relativistic physics.

Therefore, there is reason to believe that a consistent microscopic substantiation of thermodynamics within the framework of pre-relativistic physics is impossible.

\section*{Acknowledgements}

We are grateful to Prof.~Ya.I.~Granovsky, Prof.~V.V.~Uchaikin, and Dr.~V.V.~Zubkov for the stimulating discussions.
We are also sincerely grateful to Dr.~A.~Lerose for drawing our attention to the paper~\cite{Lerose}.

\end{document}